\documentclass[aps,pra,twocolumn,groupedaddress]{revtex4-2}

\usepackage{graphicx}
\usepackage{amsmath}
\usepackage{multirow}
\usepackage[T1]{fontenc}
\usepackage{xcolor}     

\begin{document}


\title{Breathing and quadrupole modes in squeezed Bose-Bose liquid drops
}


\author{Marko Korda}
\affiliation{Faculty of Science, University of Split, Ruđera Boškovića 33, HR-21000 Split, Croatia}

\author{Leandra Vranješ Markić}

\affiliation{Faculty of Science, University of Split, Ruđera Boškovića 33, HR-21000 Split, Croatia}

\author{Jordi Boronat}

\affiliation{
  Departament de Física, Universitat Politècnica de Catalunya, Campus Nord B4-B5, E-08034 Barcelona, Spain
}


\begin{abstract}
We investigate the lowest excitation modes of binary $^{39}$K ultradilute 
droplets confined 
harmonically in one direction, which reduces the 
critical number of particles needed to sustain a self-bound state. Using the 
density functional built from first-principles quantum Monte Carlo calculations 
(QMC), which incorporates the effects of finite range interactions, 
we study the crossover of the oscillation modes from three to two 
dimensions. For weak squeezing, we observe the coupling between breathing and 
quadrupole modes, resulting from the breaking of spherical symmetry. 
Increasing the squeezing, the higher-frequency mode corresponding to 
three-dimensional breathing mode becomes unstable.   For  
squeezings stronger and comparable to  experiments, we 
analyze the dependence of the excitation modes on the droplet size 
 for several interactions and show the reduction of the 
number of particles needed to sustain stable oscillations. The differences 
between the results obtained with the QMC functional and the extended 
Gross-Pitaevskii theory are increased with respect to the case of 
previously studied spherical droplets.
\end{abstract}


\maketitle

\section{Introduction}
Self-bound droplets  formed by ultracold atom mixtures have attracted 
substantial attention~\cite{bottcher_new_2020,luo_new_2021} since their 
prediction by Petrov~\cite{Petrov} and subsequent experimental detection in the 
Bose-Bose mixture of two $^{39}$K hyperfine states~\cite{Cabrera, Semeghini2018,Ferioli2019}  
and, soon after, with $^{41}$K-$^{87}$Rb~\cite{derrico_observation_2019, 
Cavicchioli_2025} and $^{23}$Na-$^{87}$Rb mixtures~\cite{Guo_2021}. The droplets 
with attractive interspecies and repulsive intraspecies interactions remain 
stable because their overall attractive mean-field energy is balanced by  
quantum fluctuations, which are in most studies  modeled by the repulsive 
Lee-Huang-Yang (LHY) term. A relevant property of these droplets is that 
they require a minimum (critical) atom number to form  a 
many-body self-bound state. There was a discernible difference between the 
critical atom number predicted by the mean field with LHY corrections 
(MF+LHY) theory and the first experiments with $^{39}$K 
mixtures~\cite{Cabrera}. This difference was reduced when finite range 
effects were taken into account, by constructing interaction potentials which 
incorporate the information on both the $s$-wave scattering length and the 
effective range.~\cite{Boronat}. 

The resulting  quantum Monte Carlo (QMC) equation of state was further used to 
construct a new density functional that enabled the study of the droplets' 
excitation modes~\cite{Leandra}. As it is well established, the excitation 
modes are stable only if their energy is below the particle emission threshold, 
determined by the absolute value of the chemical potential $\mu$~\cite{Petrov, 
Fort_2021}. Research of the lowest breathing and quadrupole modes with 
the QMC density functional showed differences of up to 20\% with those obtained 
using the extended Gross-Pitaevskii theory, which incorporates the LHY 
corrections~\cite{Leandra}. This feature indicated that finite range 
effects could also be observed by measuring the droplets' lowest excitation 
frequencies. The collective excitations of dilute droplets were also studied by 
Hu and Liu~\cite{Liu} going beyond the LHY interaction, both in free space and 
in a weak spherical harmonic potential. They observed significant changes in 
the excitation spectrum depending on the exponent $\gamma$ that appears in the 
repulsive part of the energy functional as $\propto \rho^{2+\gamma}$, with 
$\rho$ the density and $\gamma=1/2$ for the MF+LHY functional. 
However, despite significant changes in the value of $\gamma$ the excitation 
spectrum was shown to be qualitatively unchanged, consisting of bulk and surface 
modes, which for sufficiently large droplets assumed the familiar dispersion 
relations $\omega\propto k$ and $\omega \propto k^{3/2}$, respectively. On 
the other hand, the breathing modes of droplets have also recently been studied 
in imbalanced mixtures, both unconfined and 
trapped~\cite{Flynn_2023,Flynn_2024,Pelayo_2025_1}, showing additional breathing 
mode frequencies.

The critical atom number is reduced when the droplets are 
squeezed in one direction by
a harmonic confining potential~\cite{Pero}. It is therefore interesting to 
study how the oscillation frequencies change with squeezing in one direction, as 
well as with the strength of the tunable interaction, which is determined by the parameter $\delta a = a_{12}+ \sqrt{a_{11}a_{22}}$, where $a_{ij}$ are the scattering lengths.
A possible approach is to use a two-dimensional energy functional, with 
logarithmic dependence on the density~\cite{Astrakharchik}, and to connect it 
with a quasi-2D setting via $a_{\rm 2D}=2e^{-\gamma}\sqrt{\pi a_{\rm 
ho}^2/2/A}\exp[-\sqrt{\pi /2}a_{\rm ho}/a_{\rm 3D}]$, where $a_{\rm 3D}$ is the 3D 
scattering length, $a_{\rm ho}$ the harmonic oscillator length in the 
direction of confinement, $\gamma$ is the Euler's constant and $A\approx 0.905$~\cite{Petrov_2021}. Using this 
approach, breathing modes were studied in Ref. ~\cite{sturmer_breathing_2021} 
for a symmetric mixture, while Ref. ~\cite{Ma_2023} which studied dynamics of 
droplets splashing, recoiling, and deposition predicted that breathing modes can 
be probed in the deposition regime. However, 2D LHY functionals are applicable 
only for a narrow range of harmonic oscillator lengths 
$a_{\rm ho}$~\cite{Zin18,Ilg,pelayo_phases_2025,quasi-2D}. Instead, for weaker 
confinements, which are closer to experiments~\cite{Cabrera}, 3D LHY 
functionals are widely used~\cite{pelayo_phases_2025,Pelayo_2025_1}. 
The dynamical response of symmetric bosonic mixtures to an interaction quench 
in a 
quasi-2D environment, realized by tight box confinement in one direction, 
has been studied recently by Pelayo \textit{et al.}~\cite{pelayo_phases_2025}, 
for a small but positive difference of scattering lengths. Small quench 
amplitudes resulted in excitation of the breathing motion of the droplet, 
together with small amplitude bulk excitations. The same team has also explored 
the dynamical formation of droplets from imbalanced mixtures in an isotropic 
and anisotropic quasi-2D setting~\cite{Pelayo_2025_1}, observing droplet 
oscillations.  However, changes in the oscillation 
frequencies with squeezing have not yet been 
systematically explored. Also, the dependence of the modes on the finite range 
effects, which are significant in 
$^{39}$K droplets, remains unexplored.
Additionally, it has not been studied how the excitation 
modes in unconfined droplets connect to oscillation modes in two dimensions. The 
study of anisotropic trapped gases has shown coupling between monopole and 
quadrupole modes~\cite{Stringari96}, which has not been studied in ultracold 
droplets.
 
In the present work, we rely on the investigation presented in 
Ref.~\cite{Leandra}  and explore the breathing and quadrupole oscillations in a 
quasi-2D setting. In Section 2, we present the Methods used. The results are 
presented in Section 3, while Section 4 provides summary and conclusion.

\section{Methods}
We consider $^{39}$K hyperfine mixtures of $N=N_1+N_2$ atoms at the optimal 
relative atom concentration
yielded by mean field theory~\cite{Petrov} and also confirmed by quantum  
Monte Carlo simulations~\cite{Boronat},
\begin{equation}
	N_1/N_2 = \sqrt{a_{22}/a_{11}}.
	\label{optimal relative conc}
\end{equation}
The atoms are placed in an external harmonic confinement along the $z$ axis,  
$V_{\rm ext}=m\omega^2_zz^2/2$, with $\omega_z=\hbar/(m 
a_{\rm ho}^2)$. In the experiment by Cabrera \textit{et al.}~\cite{Cabrera}, 
the harmonic confinement corresponds to an oscillator length $a_{\rm 
ho}$=0.639$\,\mu$m. In order to determine the effects of squeezing, we study 
the droplets in harmonic potentials described by the oscillator 
lengths, $a_{\rm ho}=f\cdot 0.639~\mu$m.

We use density functional theory, analogously to the approach taken in Ref.~\cite{Leandra}. The energy is expressed as a function of the density 
\begin{equation}
E[\rho] = \frac{\hbar^2}{2m} \int d\mathbf{r}\, \left|\nabla \sqrt{\rho(\mathbf{r})}\right|^2
+ \int d\mathbf{r}\, (E_{\mathrm{int}}[\rho(\mathbf{r}]+V_{ext}[\rho(\mathbf{r})]).
\end{equation}
where $\Psi(\mathbf{r})$ is the effective wave function of the droplet and $\rho(\mathbf{r}) = |\Psi(\mathbf{r})|^2$ is the atom density. 
We build the many-body wave function as the product of single particle orbitals,
\begin{equation}
\Psi(\mathbf{r}_1, \mathbf{r}_2, \ldots, \mathbf{r}_N)
= \prod_{i=1}^{N_1} \psi_1(\mathbf{r}_i)
  \prod_{j=N_1+1}^{N} \psi_2(\mathbf{r}_j) \ .
\end{equation}
Minimizing the action, one obtains a Schr\"odinger like equation for each
species, 
\begin{equation}
i\hbar \frac{\partial \psi_i}{\partial t}
= \left(
- \frac{\hbar^2}{2m} \nabla^2
+ V_{\mathrm{ext}}(\mathbf{r})
+ \frac{\delta E_{\mathrm{int}}}{\delta \rho_i}
\right) \psi_i ,
\label{schro}
\end{equation}
Employing the optimal atom concentration as fixed, allows one in practice to 
solve Eq.~\eqref{schro} for the total density. 

A key step in describing quantum droplets of Bose mixtures is to establish an accurate energy functional that can be applied within the local density approximation, $\mathcal{E}_\text{int} = \rho\frac{E}{N}$.
In order to include finite-range effects, we use the QMC functional, 
developed in Ref.~\cite{Boronat}, with
\begin{equation}
	\frac{E}{N} = \alpha\rho + \beta\rho^{\gamma},
	\label{Energy per atom QMC}
\end{equation}
where $\rho$ is the total atom number density. 
The parameters $\alpha$, $\beta$, and $\gamma$ were determined~\cite{Boronat} by 
fitting the QMC energy results 
obtained using  model potentials that satisfy both the $s$-wave 
scattering length and effective range. For completeness, the parameter values, 
for the magnetic fields used in this study, are given in Table 
~\ref{tab:params}.

The same functional form has been used recently to study the stability of 
squeezed droplets~\cite{Pero}, 
allowing one to capture beyond-MF correlations encoded in QMC 
simulations. 

\begin{table}
	\centering
	\caption[Density functional parameters]{Parameters of the QMC density functional for various strengths of the magnetic field $B$, empirically determined by fitting to the DMC simulation data. These values are taken from~\cite{Pero}.}
	\label{tab:params}
    \begin{ruledtabular}
	\begin{tabular}{cccc}
		$B$ $(G)$ & $\alpha$ & $\beta$ & $\gamma$\\
		\colrule
		56.230 & -0.812 & 5.974 & 1.276 \\
		56.453  & -0.423  & 8.550 & 1.373 \\
		56.639  & -0.203 & 12.152 & 1.440 \\
	\end{tabular}
    \end{ruledtabular}
\end{table}

A similar construction can be made for the MF+LHY energy functional. For the same relative concentration, the MF+LHY energy per particle can be written as
\begin{equation}
	\frac{E/N}{|E_0|/N} = -3\left(\frac{\rho}{\rho_0}\right) + 
2\left(\frac{\rho}{\rho_0}\right)^{3/2} \ ,
	\label{Energy per atom MF-LHY}
\end{equation}
where $E_0/N$ and $\rho_0$ are the equilibrium energy per atom and equilibrium 
density,
\begin{equation}
\frac{E_0}{N}
= - \frac{25\pi^2 \hbar^2}{768 m a_{22} a_{11}}
\frac{\left( a_{12} + \sqrt{a_{11} a_{22}} \right)^3}
{\left( \sqrt{a_{11}} + \sqrt{a_{22}} \right)^6},
\end{equation}
\begin{equation}
\rho_0
= \frac{25\pi}{1024 a_{11}^3}
\frac{\left( a_{12}/a_{11} + \sqrt{a_{22}/a_{11}} \right)^2}
{\left( a_{22}/a_{11} \right)^{3/2}
\left( 1 + \sqrt{a_{22}/a_{11}} \right)^4}.
\end{equation}

In contrast to Eq.~\eqref{Energy per atom QMC}, the expression for MF-LHY energy 
per particle is written in terms of universal quantities and does not depend 
explicitly on the magnetic field $B$.

In our work,  we focus on two oscillation modes: the breathing mode and the 
quadrupole mode which correspond to the oscillations of the monopole moment 
$Q_0$ and the quadrupole moment $Q_2$, respectively.

The discrete definition of the monopole moment $Q_0$ is
\begin{equation}
	Q_0 = \sum_{i}^{N}r_i^2,
	\label{Q0 moment}
\end{equation}
where the sum is over all atoms in the system, and $r_i^2$ is the squared 
distance of particle $i$ to the center of mass of the 
system. Similarly, the quadrupole moment is defined as
\begin{equation}
	Q_2 = \sum_{i}^{N}(3x_i^2-r_i^2) = \sum_{i}^{N}(2x_i^2 - y_i^2 - z_i^2),
	\label{Q2 moment}
\end{equation}
where $x$ is selected as the axis of interest (in spherical droplets $z$ is 
usually taken).  Notice that for squeezed droplets, quadrupole modes with 
the axis of interest in the direction of squeezing or perpendicular to it are 
no longer equivalent. In our analysis, to excite the droplet we use a potential 
of the form
\begin{equation}
	V_{i} = \lambda Q_i,
	\label{Excitation potential}
\end{equation}
where $\lambda$ is a perturbative parameter of arbitrary choice, and 
$Q_i$ is the 
moment which corresponds to the oscillation mode of interest.
The choice of the perturbation potential $V_i$ ensures coupling to the desired 
collective mode. Perturbing with $V_0$ uniformly increases or decreases the 
radius of the droplet for a small amount (volume mode), while perturbing with 
$V_2$ elongates the droplet along the chosen axis (surface mode), which excites 
quadrupole oscillations. Then, after turning off the perturbation, the system 
exhibits oscillations around the equilibrium value $\langle 
\Psi(\mathbf{r}) \mid Q_i \mid \Psi(\mathbf{r}) \rangle$.
The arbitrary parameter $\lambda$ is a measure of the perturbation strength, and 
in our calculations it is kept small enough so the oscillations remain in the  
harmonic regime and linear response theory is applicable. We used $10^{-13}$ 
for $B = 56.230$ G, $10^{-14}$ for $B = 56.453$ G, and $10^{-15}$ for $B = 
56.639$.  At the end of the real time propagation, we Fourier analyzed the 
$\langle Q(t)\rangle$. The excitation modes of the droplet corresponded to the 
peaks of the obtained non-normalized strength function of the excitation 
operator $Q$.

When the majority of the multipole strength is in a single narrow peak, it is 
very useful 
to estimate the peak frequency using the sum rules approach~\cite{Stringari, 
Bohigas}.
For the monopole and quadrupole modes, the $E_1$ energies are
\begin{equation}
	E_1 (l=0) = - \sqrt{-\frac{4\hbar^2}{m}\frac{\langle r^2\rangle}{\left(\partial \langle Q_0\rangle / \partial\lambda\right)|_{\lambda=0}}},
	\label{Sum rule mono}
\end{equation}
and
\begin{equation}
	E_1 (l=2) = - \sqrt{-\frac{8\hbar^2}{m}\frac{\langle r^2\rangle}{\left(\partial \langle Q_2\rangle / \partial\lambda\right)|_{\lambda=0}}},
	\label{Sum rule quad}
\end{equation}
where $\lambda$ is the strength parameter of the perturbation in the Hamiltonian, and $\langle r^2\rangle$ is calculated for $\lambda$=0. 
In order to determine the partial derivative $\partial \langle Q_i\rangle / 
\partial\lambda$, imaginary-time  evolution is performed at several 
values of $\lambda$, typically in the range from $10^{-13}$ to 
$10^{-15}$.

The oscillations are stable when their energy is lower than the absolute 
value of the 
chemical potential $|\mu|$. 
When the energy lies above this threshold, then it is energetically favorable 
to eject atoms from the droplet, making the droplet unstable. 

We determined the dependence of $\mu=\partial E / \partial N $ on the droplet size, using the Liquid Drop Model (LDM) 
\begin{equation}
	\frac{E}{N} = E_V + E_S x + E_C x^2
	\label{eq:LDM}
\end{equation}
where $x=N^{-1/3}$, while coefficients $E_V$, $E_S$, and $E_C$,
corresponding to volume, surface, and curvature contributions to the 
energy, were determined by fitting Eq.~\eqref{eq:LDM} to the obtained data. 
For very large $N$, 
surface and curvature contributions to the chemical potential vanish, and $\mu$ 
approaches $E_V$, 
as it is expected for an infinite homogeneous system.
For small $N$, surface and curvature contributions become significant, lowering 
the absolute value of the binding energy and therefore reducing the stability of 
oscillations.

\section{Results}

\begin{figure}[]
	\centering
	\includegraphics[width=1.0\linewidth]{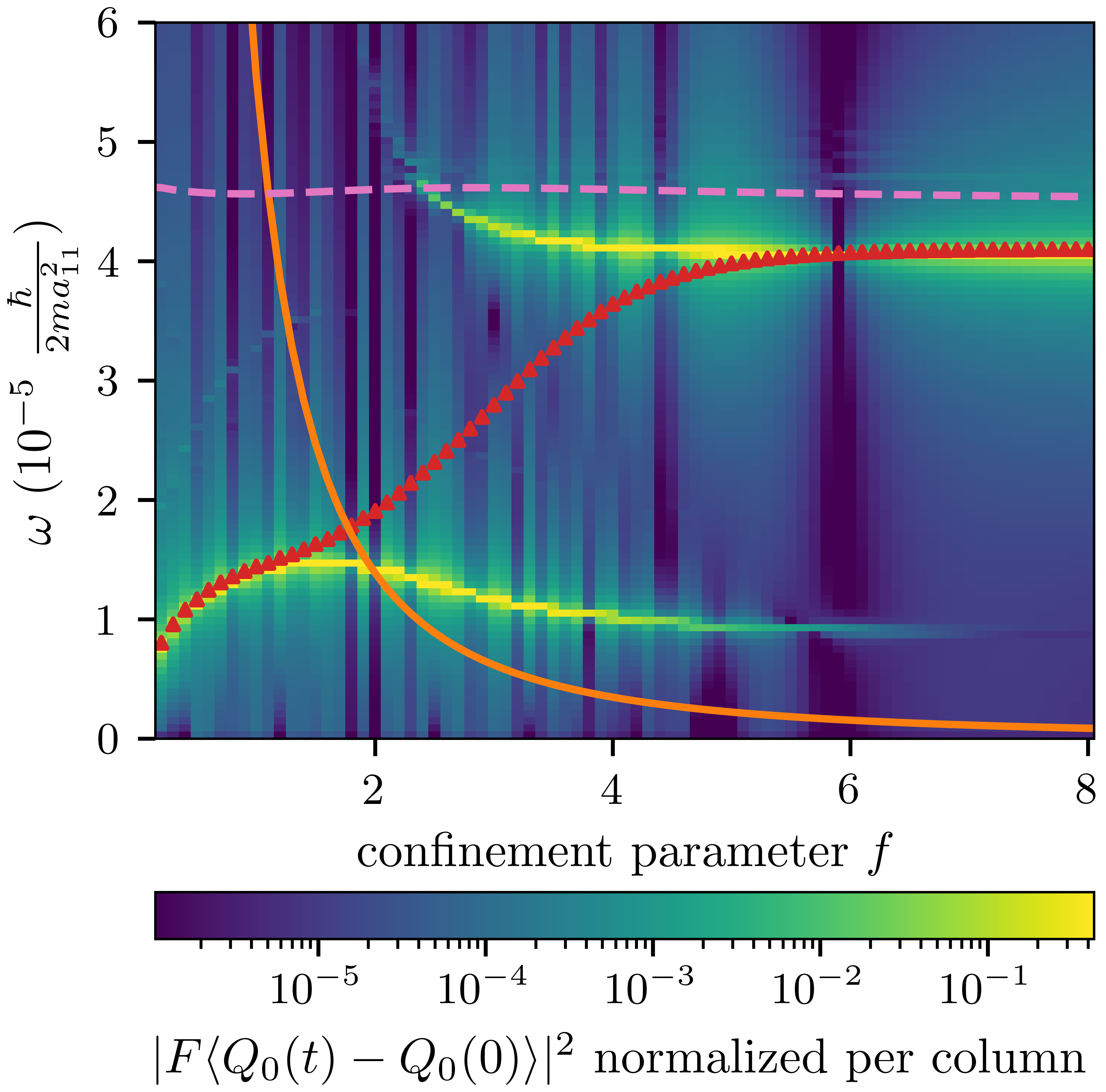}
	\caption{\label{sl:heatmap_BM} Frequency intensity distribution as a function of the squeezing parameter $f$ for $N=10^6$ droplet excited using the 3D monopole operator $V_0 = \lambda (x^2 + y^2 + z^2)$, where $\lambda = 10^{-13}$. The sum rules (red triangles) are used to approximate the weighted average frequency, $|\mu|/\hbar$ (pink dashed line) is acting as a stability threshold, and trap frequency is shown using the full orange line. The length of the harmonic oscillator used to squeeze the droplet is $a_{ho} = f \cdot 0.639$ $\mu$m, and the magnetic field is $B$=56.230 G ($\delta a=-6.53a_0$).}
\end{figure}
We squeezed the droplet along the $z$-axis using the harmonic potential defined
by its corresponding oscillator length $a_{{\rm ho}} = f a_{{\rm exp}}$, where 
$f$ is an strength factor, and $a_{{\rm exp}} = 0.639 \text{ }\mu\text{m}$ is 
the oscillator length used in the experiment of Ref.~\cite{Cabrera}. We 
investigate the range from $f=0.2$ to $f=8.0$, using the monopole perturbation 
$V_0 = \lambda (x^2 + y^2 + z^2)$, and $B$=56.230 G which corresponds to $\delta a=-6.53a_0$. The result is shown in 
Fig.~\ref{sl:heatmap_BM} for the $N=10^6$ droplet, which is large enough to have 
stable breathing oscillations in the 3D limit. We observe two dominant branches 
with a gap between them 
 in the middle region. When we increase 
$f$, the upper branch asymptotically approaches the breathing mode frequency of 
a 3D spherical droplet $(\omega_{0,3D} = 4.02(3)\cdot10^{-5}\hbar/(2ma_{11}^2))$. On the other side of the spectrum, in the lower $f$ 
range, there is a new mode that needs to be identified and which grows stronger 
as the squeezing increases. The sum rule (red triangles) is used as an 
approximation of the weighted average of the excitation spectrum, and therefore 
it gives an illustration of the crossover from one branch to another in the 
middle region. The orange line in Fig.~\ref{sl:heatmap_BM}  is the frequency of 
the squeezing harmonic trap,  showing no  
correlation between the observed oscillation frequencies and the trap frequency. 
The dashed pink line is equal to $|\mu|/\hbar$, which acts as a stability 
threshold and, as expected, one can see that the upper branch 
disappears  as it crosses that line.

In order to explain the observed excitation spectrum, we need to consider the 
symmetry of the problem.   The non-squeezed droplet possesses spherical 
symmetry, which corresponds to the $SO(3)$ symmetry group. The generators of 
that group are $L_x$, $L_y$, and $L_z$, therefore $l$ is a good quantum number. 
The monopole perturbation operator $V_0 = \lambda r^2$ is invariant under any 
rotation, therefore, it is a scalar and couples to the $l=0$ mode (breathing 
mode). The quadrupole perturbation operator $V_2 = \lambda(3x^2 - r^2)$ 
couples to the $l=2$ quadrupole mode. There is no mixing of modes for the 
spherical droplet because mixing between states with different $l$ is forbidden.

However, when we introduce an external harmonic trap along the $z$ axis, we 
break 
the $SO(3)$ symmetry, and our system now possesses the $SO(2)\times Z_2$ 
symmetry (rotational symmetry around the $z$ axis and $z$ parity symmetry). 
Consequently, the only rotational generator that remains is $L_z$, hence $m$ is 
a good quantum number. The $l=0$ state becomes $m=0$, and the $l=2$ state 
decomposes into the $m=0$, $m=\pm1$, and $m=\pm 2$ states. Therefore, in  
Fig.~\ref{sl:heatmap_BM} we observe two modes: $|l=0, m=0\rangle$ (upper 
branch) and $|l=2, m=0\rangle$ (lower branch) states. We 
can also see that the lower branch diminishes as $f$ increases because 
the droplet approaches spherical symmetry.
Also, it is important to note that the observed quadrupole mode is the axial 
$z$ quadrupole mode, while other quadrupole modes remain separated from the 
breathing mode. We do not see the $|l=1, m=0\rangle$ mode due to the parity 
symmetry $Z_2$. The states with even $l$ have positive parity, while the states 
with odd $l$ have negative parity, and only mixing between the states of the 
same parity is allowed.

\begin{figure}[]
	\centering
	\includegraphics[width=1.0\linewidth]{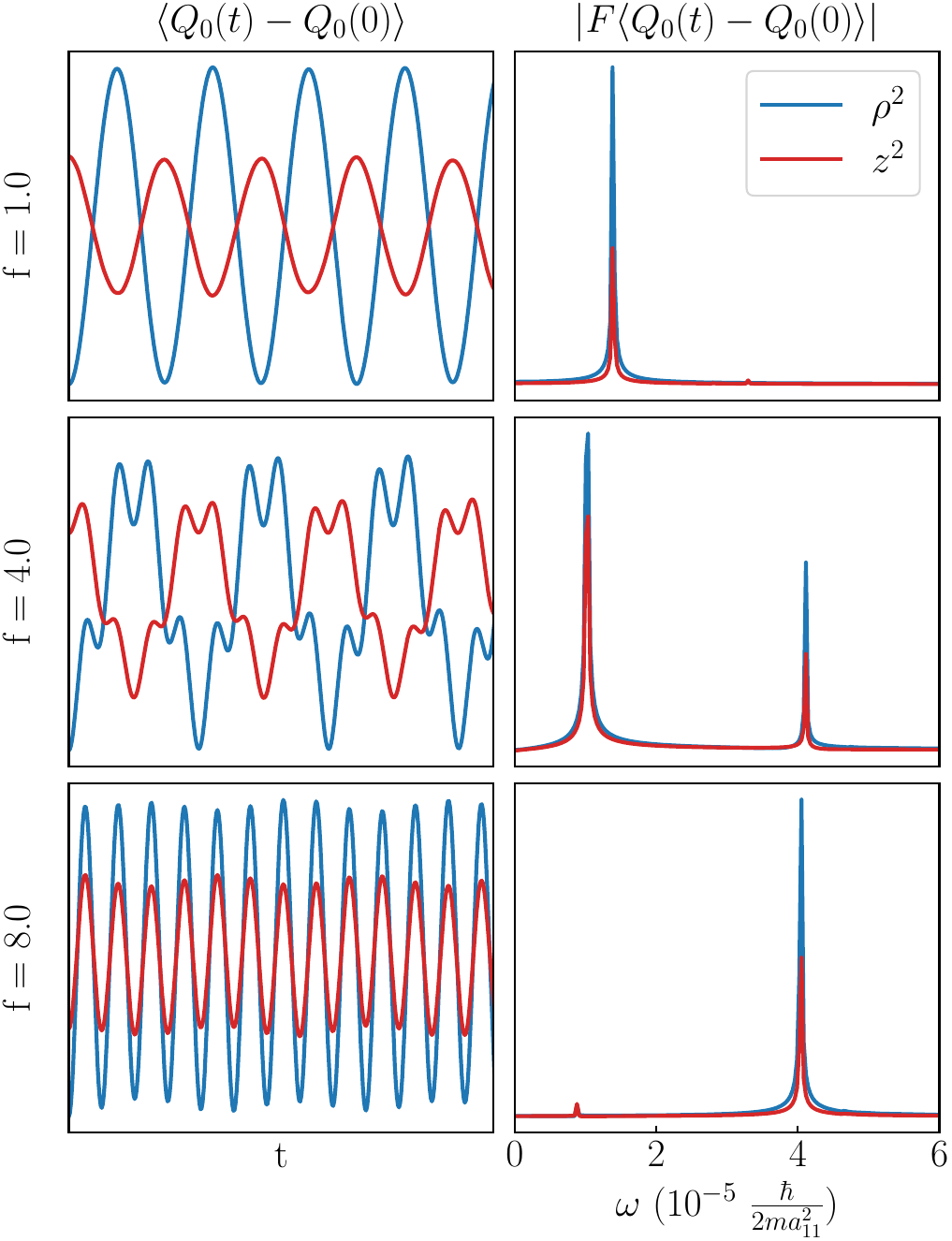}
	\caption{\label{sl:phases} Comparison between $\rho^2$ and $z^2$ oscillations (left figures), and comparison between $\rho^2$ and $z^2$ frequency amplitudes (right figures) for different values of the squeezing parameter $f$, for the droplet in Fig.\ref{sl:heatmap_BM}.}
\end{figure}

To better visualize the character of the modes, we compare the oscillations of 
$\langle \rho^2(t)\rangle$
and $\langle z^2(t)\rangle$ for three values of $f$ in Fig.~\ref{sl:phases}. 
For $f=1.0$ we have only the lower frequency, and we can see that $\langle 
\rho^2(t)\rangle$ and $\langle z^2(t)\rangle$ oscillate exactly in antiphase, 
which is the property of axial quadrupole mode. However, it is important to note 
that both the oscillation and frequency amplitudes of $\langle z^2(t)\rangle$ 
for $f=1.0$ are magnified 50 times to show on the same scale as the $\langle 
\rho^2(t)\rangle$ amplitudes. Consequently, when we approach the 
quasi-two-dimensional regime, this mode can be renamed the 2D radial breathing 
mode, as the oscillations in $z^2$ are practically negligible. On the other side 
of the spectrum, for $f=8.0$ we have only the higher frequency, and we can see 
that $\langle \rho^2(t)\rangle$ and $\langle z^2(t)\rangle$ oscillate in phase, 
which is the property of the spherical breathing mode.  For $f=4.0$ we observe 
the superposition of both modes.

\begin{figure}[]
	\centering
	\includegraphics[width=1.0\linewidth]{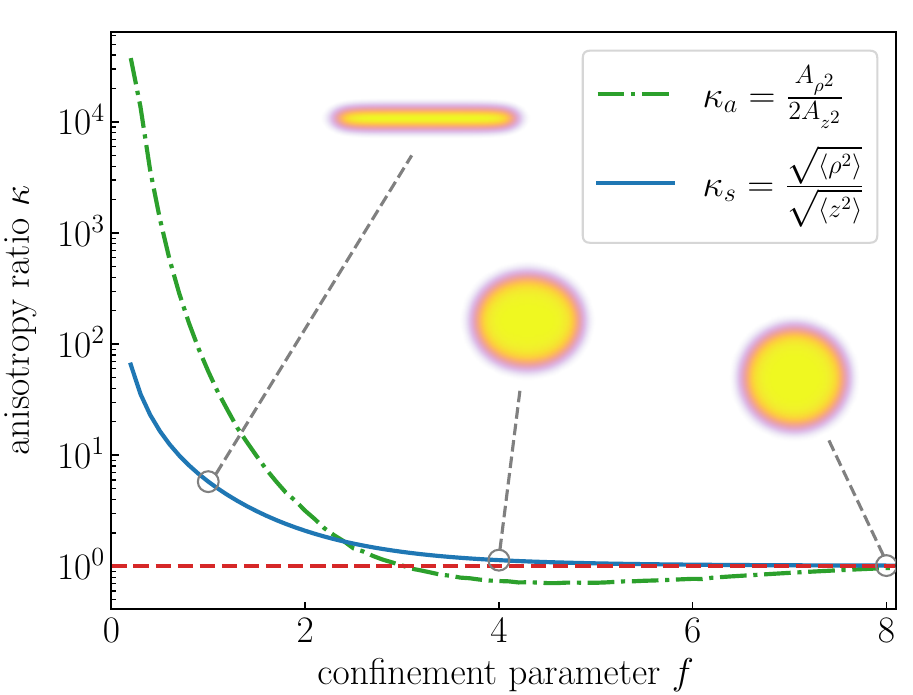}
	\caption{\label{sl:anisotropy} Spacial anisotropy (blue line) and oscillation anisotropy (green line) as a function of the squeezing parameter $f$, for the droplet in Fig.\ref{sl:heatmap_BM}. The red dashed line at $\kappa = 1$ is an anisotropy ratio of the spherical droplet as the reference. The droplets for $f=1.0$, $f=4.0$, and $f=8.0$ are drawn to illustrate the squeezing strength.}
\end{figure}

Further, we analyze the anisotropy ratios of spatial widths and of oscillation 
amplitudes for the entire squeezing range. We define the spatial anisotropy 
ratio $\kappa_s=\sqrt{\langle \rho^2 \rangle}/\sqrt{\langle z^2 \rangle} $, 
and the 
oscillation anisotropy ratio $\kappa_a= A_{\rho^2}/(2 A_{z^2})$, with  
$A_{\rho^2}$ and $A_{z^2}$ the amplitudes of $\langle \rho^2(t)\rangle$ and 
$\langle z^2(t)\rangle$, respectively. Notice that $A_{\rho^2}/A_{z^2}$ takes a 
value of 2 for the spherical droplet, therefore, we further divide 
it by 2 in $\kappa_a$ for visual consistency. The results are shown in 
Fig.~\ref{sl:anisotropy}. We can see that for values of $f\le 1.0$, the radial 
oscillations are 2 to 4 orders of magnitude stronger than the axial 
oscillations.
In this case, the droplet is significantly squeezed and approaches the 
quasi-two-dimensional 
regime. Instead, for $f\ge 3.0$ we see that the axial oscillations become 
stronger than the radial oscillations because the higher mode becomes 
significant and the two modes add up for a fraction of the period. This droplet 
is already very weakly squeezed with spatial anisotropy close to 1. Increasing 
$f$ further drives both anisotropy values to 1, which corresponds to 
the spherical droplet.

\begin{figure}[]
	\centering
	\includegraphics[width=1.0\linewidth]{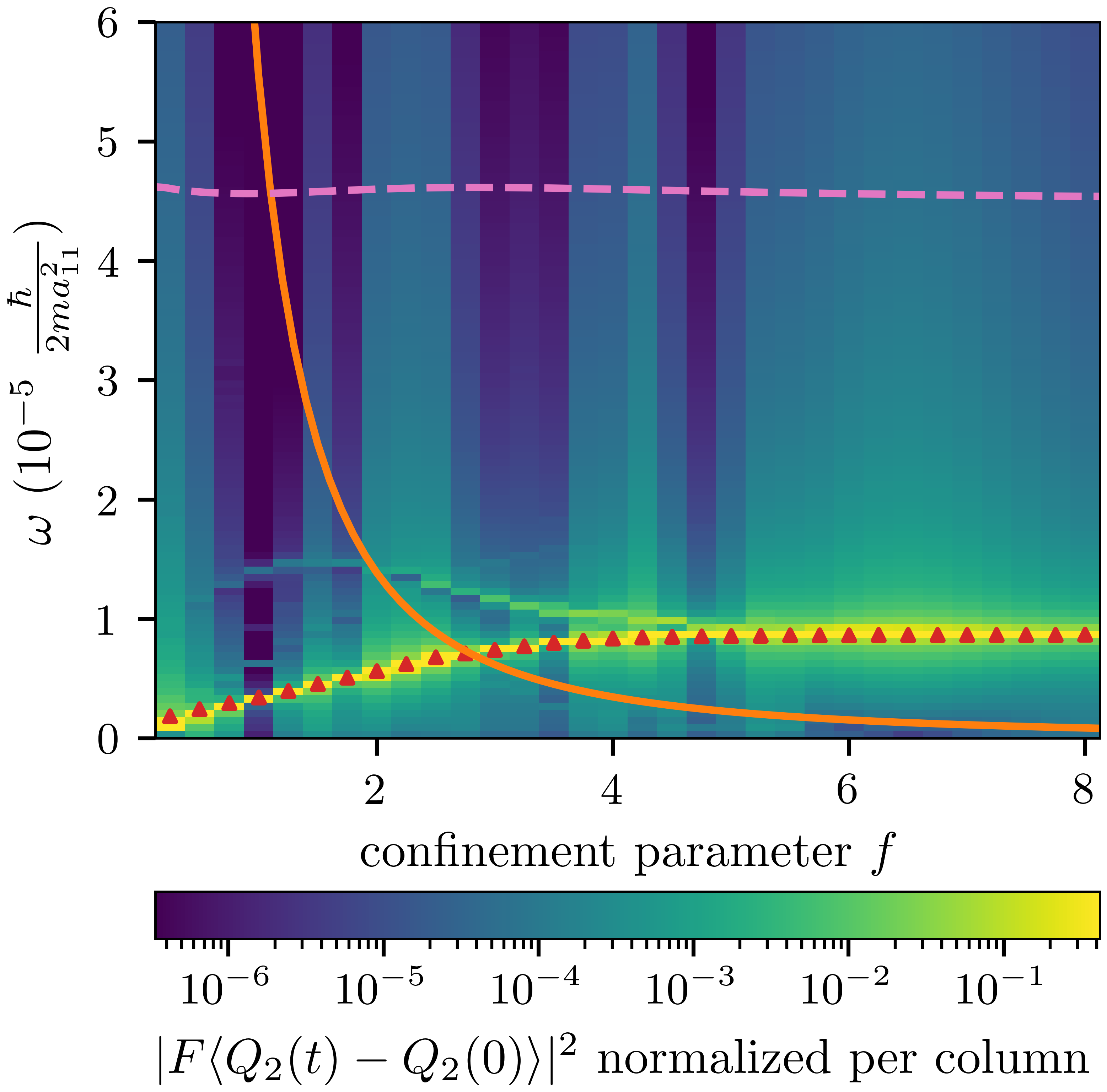}
	\caption{\label{sl:heatmap_QM} Frequency intensity distribution as a function of the squeezing parameter $f$ for $N=10^6$ droplet excited using the 3D quadrupole operator $V_2 = \lambda (2x^2 - y^2 - z^2)$, where $\lambda = 10^{-13}$. The sum rules (red triangles) are used to approximate the weighted average frequency,  $|\mu|/\hbar$ (pink dashed line) is acting as a stability threshold, and trap frequency is shown using the full orange line. The length of the harmonic oscillator used to squeeze the droplet is $a_{ho} = f \cdot 0.639$ $\mu$m, and the magnetic field is $B$=56.230 G ($\delta a=-6.53a_0$).}
\end{figure}

We also performed a similar analysis using the $V_2 = \lambda(3x^2 - 
r^2)$, which should 
excite a separate quadrupole mode, and the results are shown in 
Fig.~\ref{sl:heatmap_QM}. Now, one can see only one dominant mode as 
expected, the sum rules showing exceptional accuracy. But we also see some sort 
of shadow that resembles the shape of the lower branch from 
Fig.~\ref{sl:heatmap_BM}. This is because we used the full three-dimensional 
quadrupole operator which has a weak but nonzero coupling to the $|l=2, 
m=0\rangle$ state. The same analysis was repeated using the two-dimensional 
quadrupole operator $V_{2,2D} = \lambda (x^2 - y^2)$ and the shadow disappeared, 
while the dominant branch remained unchanged.  Also, it is important 
to note that this branch and the lower one in Fig.~\ref{sl:heatmap_BM} both 
approach the same value when increasing $f$, which is the quadrupole frequency 
of the spherical droplet $(\omega_{2,3D} = 8.8(3)\cdot10^{-6}\hbar/(2ma_{11}^2))$, because in the spherical limit all quadrupole modes are equivalent.

To study the dependence of the oscillations on the size of the 
droplet 
and the interaction strength, we have explored  squeezing 
parameters $f\le 1$ ($a_{{\rm ho}}\le0.639$ $\mu$m), for which  the critical atom 
number was shown to be 
lower~\cite{Cabrera,Pero} than in 3D~\cite{Semeghini2018,Semeghini}. From the previous 
analysis, we expect in this regime a single monopole mode (to 
be in fact the axial quadrupole mode) which we will  call 2D 
breathing mode. Instead, when referring to the quadrupole mode, we will 
consider the mode with the relevant axis perpendicular ($x$) to the squeezing 
direction ($z$).

\begin{figure}[]
	\centering
	\includegraphics[width=1.0\linewidth]{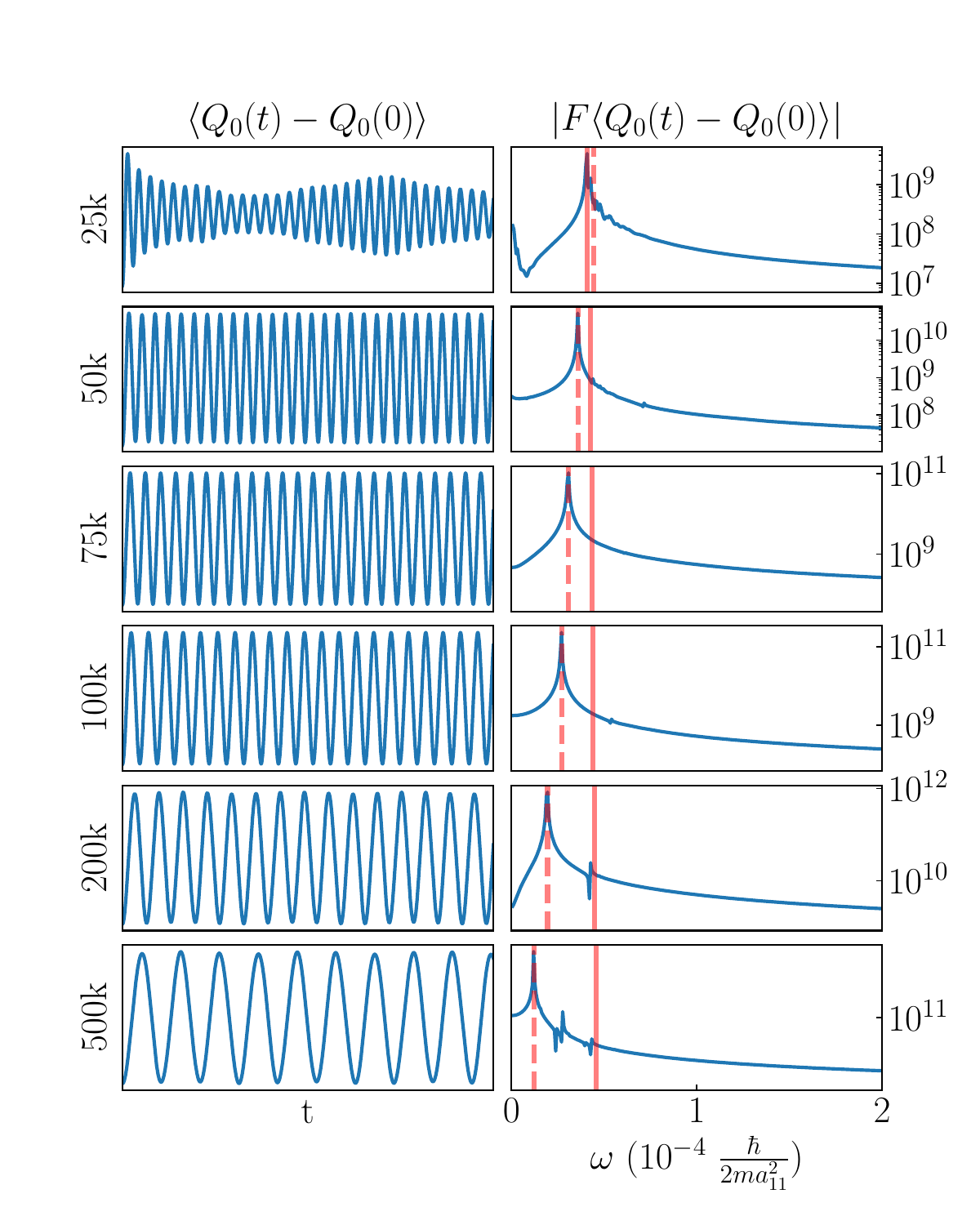}
	\caption{\label{sl:Qt 1e-13 0.25 BM} Time evolution of the monopole moment oscillations $\langle Q_0(t)\rangle$ (left) obtained by the QMC energy functional, and the corresponding frequency spectrum (right) obtained by the Fourier transform. In the right figures, the vertical red dashed lines correspond to the frequency $E_1/\hbar$ obtained by the sum rule in Eq.~\eqref{Sum rule mono}, and the vertical red solid lines correspond to the frequency $|\mu|/\hbar$. The squeezing parameter is $f=0.25$, and $B$=56.230 G ($\delta a=-6.53a_0$).}
\end{figure}

Fig.~\ref{sl:Qt 1e-13 0.25 BM} on the left shows $\langle Q_0(t)\rangle$ for 
droplets of an increasing number of atoms with $f = 0.25$, while the plots on 
the right hand side show the monopole strength function on a logarithmic scale 
as a function of the excitation frequency.
The dashed vertical lines correspond to the frequency $E_1/\hbar$ obtained by
the sum rule (Eq.~\eqref{Sum rule mono}) which shows exceptional agreement with the 
peaks obtained by the calculations.
This aligns with the expectations, as the sum rules approximate the weighted average frequency in the spectrum and, for all values of $f$, we can see a very dominant peak.
Vertical solid line shows $|\mu|/\hbar$, where the chemical potential $\mu$ 
represents the atom 
emission threshold, i.e., the oscillation energy required for atoms to be 
ejected from the droplet.  For $f = 0.25$, the oscillations are clearly harmonic 
and stable for droplets with $N = 50000$ or more.
Only the droplet with $N = 25000$ exhibits a beating behavior, pointing to the  
instability due to the peak frequency approaching the stability 
threshold.
Decreasing the squeezing, the beating behavior shifts to higher $N$, e.g. for $f = 1.0$, it is observed for $N = 50000$ and $N = 75000$ droplets, with only $N = 100000$ being stable and harmonic. In all cases, it can be explained by the relation between the predicted oscillation frequency and $|\mu|/\hbar$ as the stability threshold. $\mu$ depends on the number of particles, but the variations are simply too small to be appreciable on the scale of the x-axis. 

\begin{figure}[]
	\centering
    \includegraphics[width=1.0\linewidth]{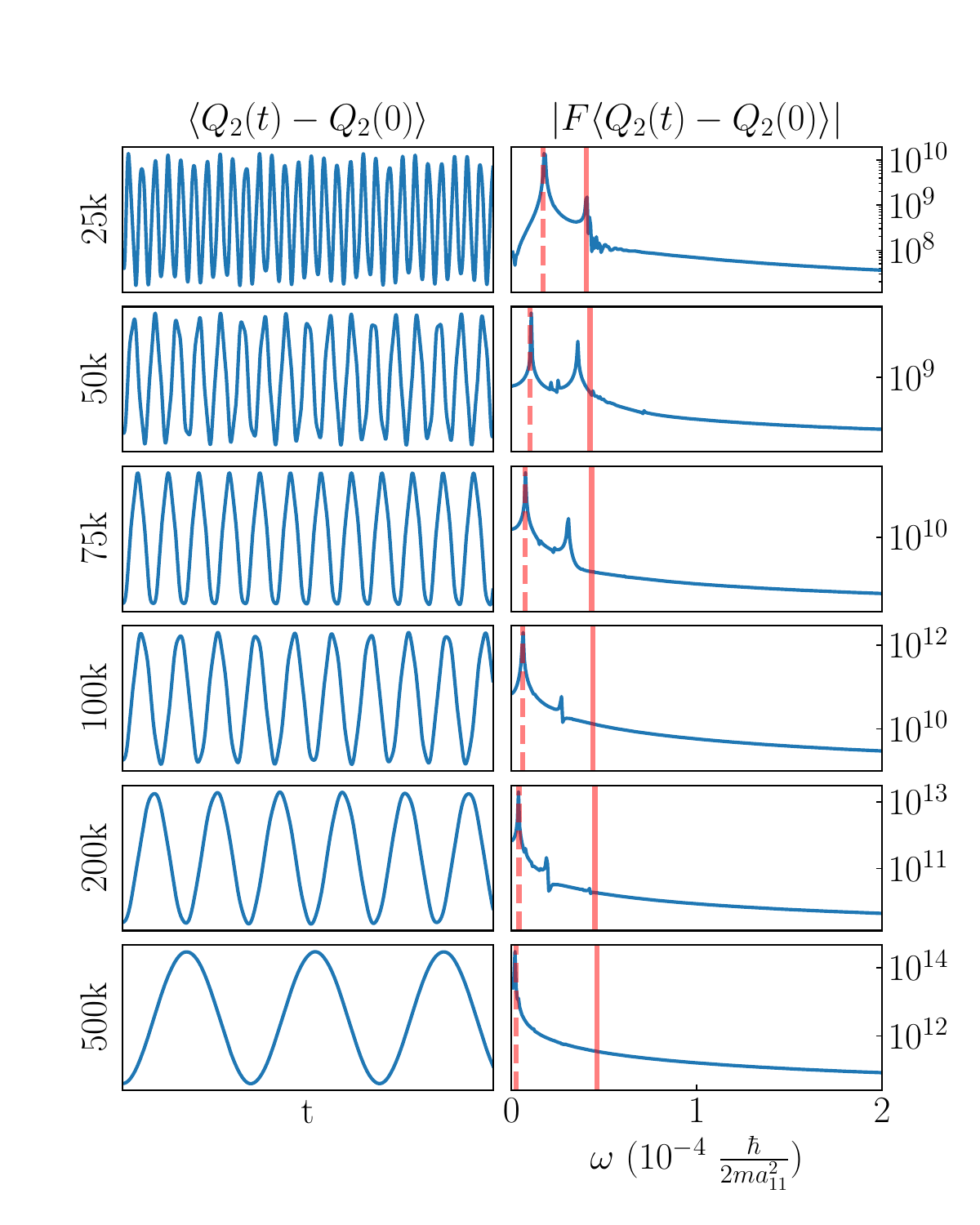}
	\caption{\label{sl:Qt 1e-13 0.25 QM} Time evolution of the quadrupole moment oscillations $\langle Q_2(t)\rangle$ (left) obtained by the QMC energy functional, and the corresponding frequency spectrum (right) obtained by the Fourier transform. In the right figures, the vertical red dashed lines correspond to the frequency $E_1/\hbar$ obtained by the sum rule in Eq.~\eqref{Sum rule quad}, and the vertical red solid lines correspond to the frequency $|\mu|/\hbar$. The squeezing parameter is $f=0.25$, and $B$=56.230 G ($\delta a=-6.53a_0$).}
\end{figure}

Fig.~\ref{sl:Qt 1e-13 0.25 QM} shows the same information as Fig.~\ref{sl:Qt 
1e-13 0.25 BM}, 
but for a quadrupole mode (Eq.~\eqref{Q2 moment}). Again, we observe excellent 
agreement between the frequency $E_1/\hbar$ obtained by the sum rule (Eq.~\eqref{Sum rule quad}) and the peaks calculated with DFT.
For the smallest $f$, the oscillations seem to be stable and mostly harmonic, 
but we 
can notice a smaller second peak showing at higher frequency, just below the 
stability threshold.
Instead, larger droplets show a single peak frequency, which dominates 
other possible peaks. 
As an example, the oscillation energy for the droplet with $N = 500000$ is completely concentrated in a single narrow peak. Decreasing the squeezing to $f=1.0$, the frequency spectrum is likewise dominated by a single peak.
Comparing to monopole oscillations, the quadrupole oscillations happening at 
lower frequencies 
tend to be more stable throughout the range of $f$ and $N$; only the $N = 
25000$ atom droplet becomes unstable for the squeezing strength $f = 1.0$. 
Therefore, strongly confined droplets can exhibit stable 
quadrupole 
collective oscillations for very small atom numbers.  

The dependence of the oscillation frequencies on the number of atoms in 
the droplet, 
for squeezings in a range from $f=0.25$ to $f=1.0$ is shown in Fig.~\ref{sl:f 
comp Both}. Each panel shows a different magnetic field which determines the 
interaction strength, measured in terms of  $\delta 
a=a_{12}+\sqrt{a_{11}a_{22}}$. For both modes, we can see that increasing the 
squeezing reduces the peak frequency throughout the $N$ range and 
that the 2D breathing frequencies are higher than the quadrupole ones (as it 
happens in spherical droplets). 
In all panels, the chemical potential, determined using the liquid drop model (Eq.~\eqref{eq:LDM})
for each squeezing, is shown as well. 2D breathing mode frequencies cross the 
$|\mu|/\hbar$ line sooner, demonstrating that in the entire range of squeezings, 
the critical number of atoms needed to support stable oscillations is 
systematically lower for quadrupole modes.

Finally, we compare results for different magnetic fields. Besides 
$B$=56.230 G 
($\delta a=-6.53a_0$), the calculations were additionally performed at fields 
$B$=56.453 G($\delta a=-4.41a_0$), and $B$=56.639 G($\delta a=-2.44a_0$), 
corresponding to weaker interactions. Similarly to what was observed in  
unconfined droplets~\cite{Leandra}, excitation mode frequencies are lower for 
weaker interactions. Furthermore, the size of the droplets required to sustain 
stable oscillations increases. The sum rule remains in very good agreement for 
all magnetic fields for the chosen squeezing range,  a signal of a single 
dominant oscillation peak. The only exception is somehow the small disagreement 
in breathing mode frequencies for $f=1.0$ and $B=56.230$ G, which 
could signal the onset of the 3D breathing mode.


For the magnetic field $B=56.230$ G we have also performed the calculations 
using the MF+LHY functional and obtained qualitative agreement. Since the 
stability of the oscillations depends on the chemical potential, we have 
investigated  how the LDM coefficients depend on the value of $f$, since each 
coefficient has a direct physical interpretation.
The corresponding coefficients  ($E_V$, $E_S$, $E_C$) are 
reported in Table~\ref{tab:LDM SQ} for both QMC and  MF+LHY, 
and for every $f$ value.
As we can see, the coefficients from the QMC simulations tend to have greater absolute value than the MF+LHY counterparts, keeping the same sign.

\begin{figure}[]
	\centering
	\includegraphics[width=1.0\linewidth]{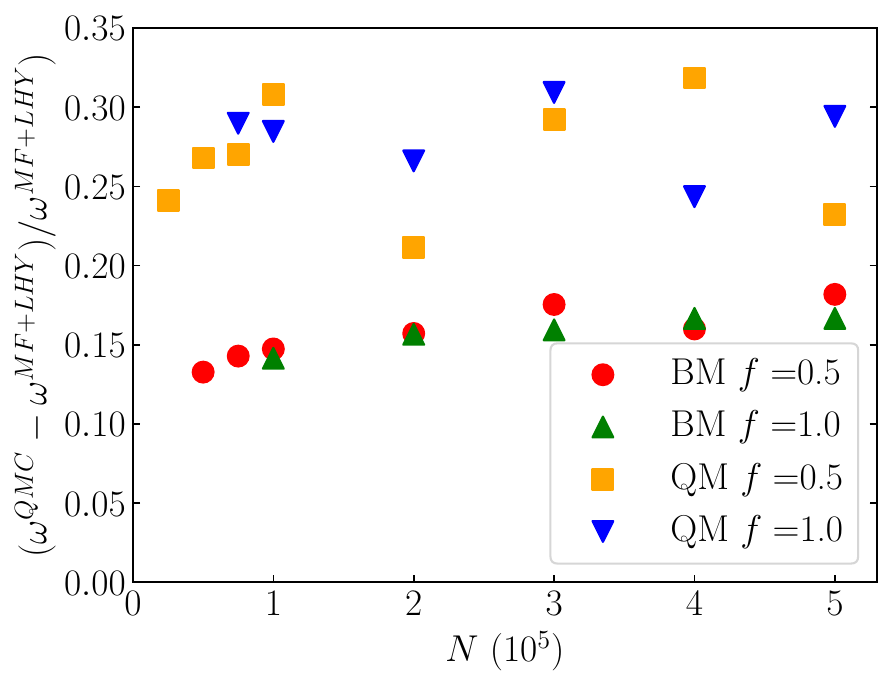}
	\caption{\label{sl:f100 error} Comparison between the results for 2D breathing (BM) and quadrupole (QM) mode obtained using QMC and MF+LHY functionals, for squeezing strength parameters $f=0.5$ and $f=1.0$.}
\end{figure}

As for oscillation frequencies, we find that MF+LHY generally predicts lower 
frequencies (see Fig.~\ref{sl:f100 error}). For the 2D breathing mode the 
values are $13\% - 20\%$ lower, and for the quadrupole mode the difference is up 
to about $30\%$,  which is larger than for 3D droplets~\cite{Leandra}, where the MF+LHY 3D breathing  and quadrupole mode frequencies for this magnetic field were about $12\%$  and $20\%$ lower, respectively.

{
\renewcommand{\arraystretch}{1.2}
\begin{table}
	\centering
	\caption[$\mu$ for different $f$ values]{Liquid drop model coefficients obtained by fitting the Eq.~\eqref{eq:LDM} on results from both the QMC and MF+LHY simulations, for different values of the squeezing strength parameter $f$ and $B$=56.230 G. All coefficients share the same dimension with the energy.}
	\label{tab:LDM SQ}
    \begin{ruledtabular}
	\begin{tabular}{ccccc}
		 $f$ & Functional & $E_V$ ($10^{-5}$)& $E_S$ ($10^{-4}$)& $E_C$ ($10^{-3}$)\\
        \colrule
		\multirow{2}{*}{$0.25$} & QMC & $-4.860$ & $2.726$ & $5.235$ \\
		 & MF+LHY & $-4.190$ & $2.057$ & $4.571$ \\
		\multirow{2}{*}{$0.50$} & QMC & $-5.194$ & $6.176$ & $3.305$ \\
		 & MF+LHY & $-4.408$ & $4.360$ & $4.020$ \\
		\multirow{2}{*}{$0.75$} & QMC & $-5.601$ & $10.213$ & $-1.142$ \\
		 & MF+LHY & $-4.717$ & $7.371$ & $1.090$ \\
		\multirow{2}{*}{$1.00$} & QMC & $-5.989$ & $14.073$ & $-6.182$ \\
		 & MF+LHY & $-5.066$ & $10.735$ & $-3.203$ \\
	\end{tabular}
    \end{ruledtabular}
\end{table}
}

\begin{figure}[]
	\centering
	\includegraphics[width=1.0\linewidth]{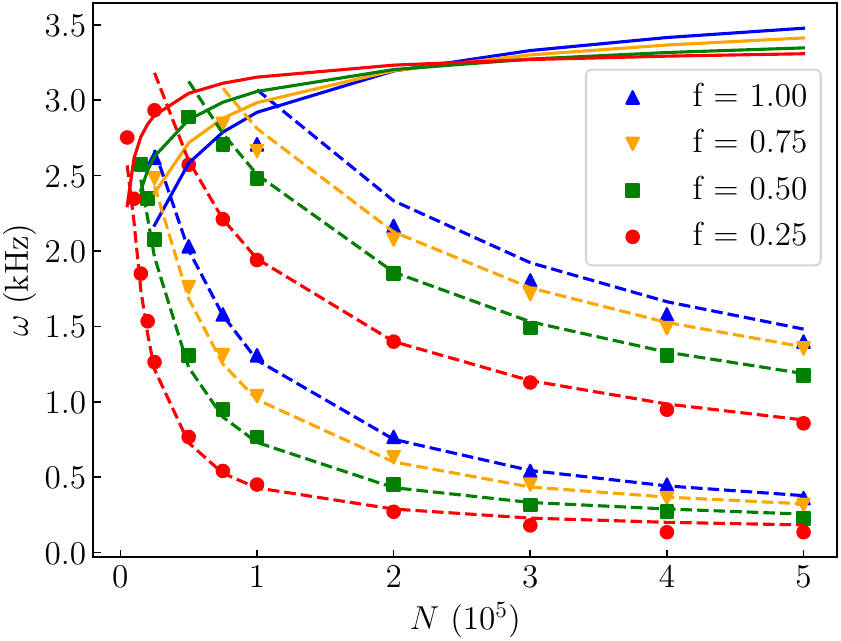}
    
    \vspace*{0.2cm}
    
    \includegraphics[width=1.0\linewidth]{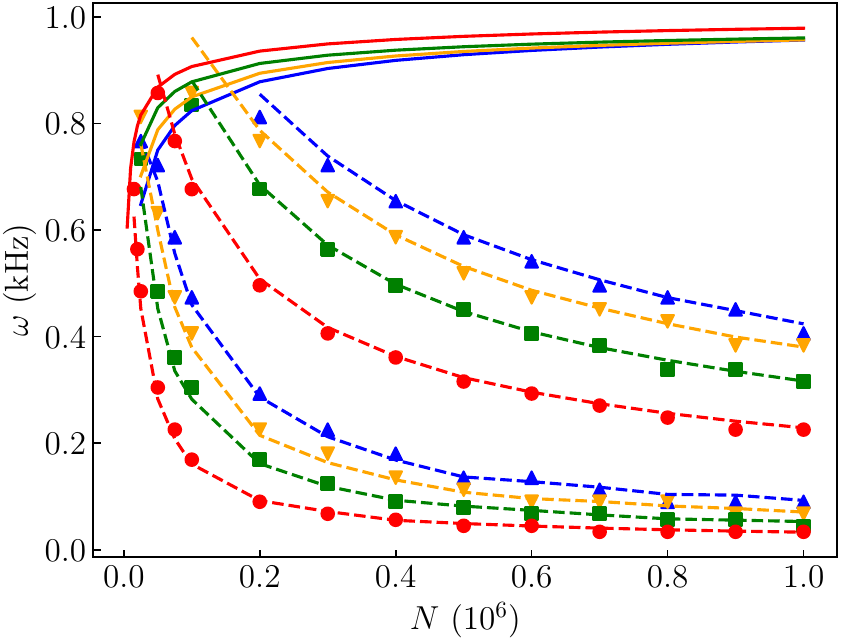}
    
    \vspace*{0.2cm}
    
    \includegraphics[width=1.0\linewidth]{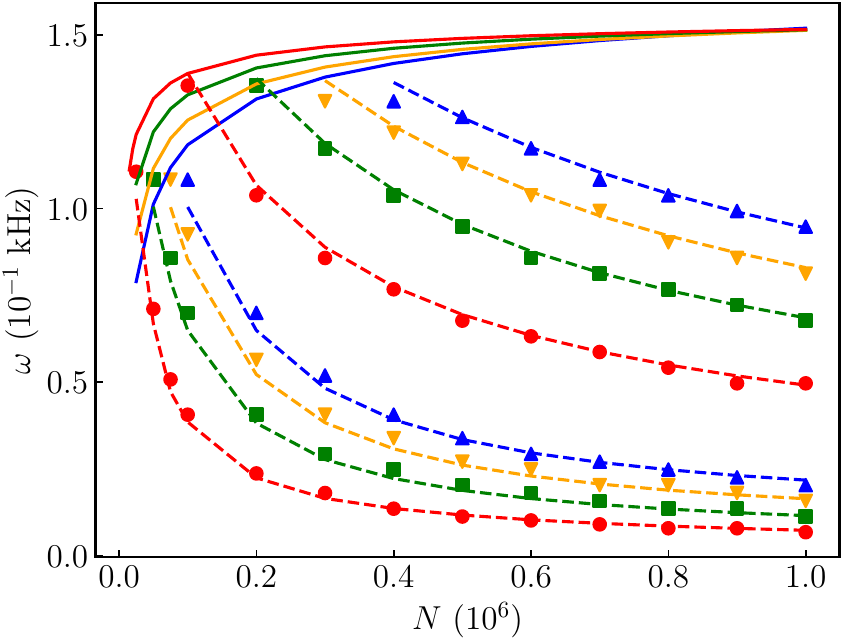}
    
	\caption{\label{sl:f comp Both} 2D breathing mode (upper) and quadrupole mode (lower) frequencies as a function of the number of atoms $N$ compared between various squeezing strength parameters $f$, obtained by the QMC functional simulations. Dashed lines correspond to the $E_1/\hbar$ frequencies from the sum rule (Eq.~\eqref{Sum rule mono}). Solid lines represent the $|\mu|/\hbar$ frequencies. Top figure corresponds to $B$=56.230 G ($\delta a=-6.53a_0$), middle to  $B$=56.453 G($\delta a=-4.41a_0$), and bottom to $B$=56.639 G($\delta a=-2.44a_0$).}
\end{figure}

\section{Summary and conclusion}

A study of the breathing and quadrupole modes of hyperfine $^{39}$K Bose-Bose 
droplets was performed 
as a function of the droplet confinement in one direction. For a large droplet 
we observe significant changes in oscillation modes depending on the 
dimensionality. In the harmonic confinement in the $z$ direction, the droplet 
loses the spherical symmetry, resulting in a coupling between the breathing and 
axial quadrupole modes. Increasing the squeezing, the breathing mode 
frequency increases and eventually disappears when its energy becomes higher 
than the chemical potential. On the other hand, with increased squeezing, the 
lower branch mode, corresponding to axial quadrupole oscillations with $m$=0, 
shows weaker oscillations in the $z$ direction and transitions to effectively 
a 2D breathing mode. The coupling between modes has previously been predicted 
in trapped Bose and Fermi 
systems,~\cite{Stringari96,heiselberg_collective_2004,maruyama_coupled_2007,
maruyama_deformation_2014} but, to the best of our knowledge, it has not been 
discussed for ultracold droplets.

We have further shown that squeezing reduces the minimum number of atoms 
needed to sustain 
stable excitation modes. With an increase in the number of atoms, we observe 
the reduction of the frequency, which generally  decreases with the increase in 
the cluster size.  This behavior therefore differs from the one in spherical 
harmonic confinement, where non monotonic dependence of the breathing mode 
frequency on the number of particles was reported for the same 
mixture~\cite{Liu}, which is a consequence of the constraint on the size of the 
droplet. Weaker interactions also result in lower frequencies. However, except 
for the weakest interactions, the timescale of the oscillations is of the order 
of a few milliseconds, which suggests that they could be detected in experiments, 
where lifetimes of the order of 10 ms were reported for squeezed droplet 
corresponding to $f=1.0$~\cite{Cabrera}. In order to test this assumption, one 
can include three-body losses (3BL) in the real-time evolution of the 
wave function by adding the term $-i\hbar K_{111}\rho_1^2/2 $ to  
Eq.~\eqref{schro}, neglecting the very small 3BL in other 
channels~\cite{Semeghini}. For a rough estimate, we instead did several 
exploratory breathing mode calculations for $f=0.25$, modeling 3BL with the 
total density. It is worth noticing that there is a large uncertainty in the 
experimentally determined values of $K_{111}$, but since in our calculations 3BL 
were acting on both components, we chose the lowest reported values of 4.5 
$\times 10^{-28}$ cm$^6$/s. With that value and for non-oscillating smaller 
droplets, which do not support stable oscillations, we could reproduce the 
lifetime reported in Ref.~\cite{Cabrera}. When increasing the number of atoms, 
we observed a much more rapid decay than the one reported in 
Ref.~\cite{Cabrera}, so we were able to observe only one full damped 
oscillation in a droplet of $2\times 10^{5}$ atoms for any $B$. Despite that, 
we could recover a very broad peak at the correct frequency. The reason
behind the reduction of the 
droplet lifetime in larger saturated droplets is that they have larger 
density, which increases substantially the droplet decay, $\partial /\partial t 
\int \rho dV =-K_{111}\int \rho^3 dV$. Decreasing  $K_{111}$ below the 
experimental values, more oscillations were observed, with periods decreasing 
in accordance with the loss of atoms. A more extensive study, without the fixed 
ratio of $N_1/N_2$ and the 3BL acting only in the first component, is needed to 
fully assess the most optimal conditions for observing the droplet oscillations.

The effects observed in the present work 
are expected to also occur for heteronuclear droplets, which have the same
binding mechanism, but importantly can  reach lower central 
densities~\cite{derrico_observation_2019,Cikojevic_KRb}. Particularly appealing 
are  K-Rb 
droplets, which have a significantly  longer lifetime than $^{39}$K droplets, on 
the order of 100 ms, 
due to lower densities and lower 3BL, which act mainly in the K-Rb-Rb channel 
with $K=(7\pm 4)\times 10^{-29}$ cm$^6$/s~\cite{derrico_observation_2019}. They 
were studied in 3D, using both QMC and MF+LHY functionals,~\cite{Cikojevic_KRb} 
that can be easily applied here. Our research in progress for these droplets 
shows the lowering of the critical number of atoms in K-Rb droplets with 
squeezing. Therefore, we expect that stable oscillations could be observed 
for experimentally achievable droplet sizes, potentially demonstrating the 
coupling between different modes. The recent observation of breathing-like 
oscillations of K-Rb droplets in the waveguide, achieved by suddenly changing 
interspecies attraction~\cite{Cavicchioli_2025}, supports this expectation. The 
same experiment has also observed that droplets excited in the 
compression-elongation mode, for large elongations, undergo a capillary 
instability and fragment into multiple smaller droplets, which is an effect that 
would be interesting to study in other geometries.

We have worked out with the QMC density functional, which incorporates 
information on $s$-wave scattering lengths 
and effective ranges, thus going beyond universal LHY effects. In selected 
cases the results were compared with the ones obtained using MF+LHY functional. 
Similarly to previous work~\cite{Leandra}, the QMC functional predicts higher frequencies than the MF+LHY functional, with the difference between the two predictions increasing up to 
30\%. It is important to note 
that both functionals are fully three-dimensional, as the use of strictly 2D 
functionals would lead to significantly different predictions of droplet 
properties~\cite{quasi-2D}, including 
oscillation modes.~\cite{pelayo_phases_2025}


\begin{acknowledgments}
We thank Ivan Poparić for helpful discussions.
We acknowledge financial support from the Croatian Science Foundation projects 
IP-2022-10-6144  and  DOK-2025-03-3299 and from the 
Spanish Ministerio de Ciencia, Innovación y Universidades (grant 
PID2023-147469NB-C21, financed by MICIU/AEI/10.13039/501100011033 and FEDER-EU).
The computational resources of UniST-Phy server at the University of Split and Padobran cluster at SRCE in Zagreb were used.
\end{acknowledgments}

\bibliography{References}

@article{Petrov,
  title = {Quantum Mechanical Stabilization of a Collapsing {B}ose-{B}ose Mixture},
  author = {Petrov, D. S.},
  journal = {Phys. Rev. Lett.},
  volume = {115},
  issue = {15},
  pages = {155302},
  numpages = {5},
  year = {2015},
  month = {Oct},
  publisher = {American Physical Society},
  doi = {10.1103/PhysRevLett.115.155302},
  url = {https://link.aps.org/doi/10.1103/PhysRevLett.115.155302}
}

@article{Cabrera,
    author = {C. R. Cabrera  and L. Tanzi  and J. Sanz  and B. Naylor  and P. Thomas  and P. Cheiney  and L. Tarruell },
    title = {Quantum liquid droplets in a mixture of {B}ose-{E}instein condensates},
    journal = {Science},
    volume = {359},
    number = {6373},
    pages = {301-304},
    year = {2018},
    doi = {10.1126/science.aao5686},
}

@article{Semeghini2018,
  title = {Self-Bound Quantum Droplets of Atomic Mixtures in Free Space},
  author = {Semeghini, G. and Ferioli, G. and Masi, L. and Mazzinghi, C. and Wolswijk, L. and Minardi, F. and Modugno, M. and Modugno, G. and Inguscio, M. and Fattori, M.},
  journal = {Phys. Rev. Lett.},
  volume = {120},
  issue = {23},
  pages = {235301},
  numpages = {5},
  year = {2018},
  month = {Jun},
  publisher = {American Physical Society},
  DOI = {10.1103/PhysRevLett.120.235301},
  url = {https://link.aps.org/doi/10.1103/PhysRevLett.120.235301}
}

@article{Ferioli2019,
  title = {Collisions of Self-Bound Quantum Droplets},
  author = {Ferioli, Giovanni and Semeghini, Giulia and Masi, Leonardo and Giusti, Giovanni and Modugno, Giovanni and Inguscio, Massimo and Gallem{\'{\i}}, Albert and Recati, Alessio and Fattori, Marco},
  journal = {Phys. Rev. Lett.},
  volume = {122},
  issue = {9},
  pages = {090401},
  numpages = {5},
  year = {2019},
  month = {Mar},
  publisher = {American Physical Society},
  doi = {10.1103/PhysRevLett.122.090401},
  url = {https://link.aps.org/doi/10.1103/PhysRevLett.122.090401}
}

@article{Semeghini,
  title = {Dynamical formation of quantum droplets in a $^{39}\mathrm{K}$ mixture},
  author = {Ferioli, G. and Semeghini, G. and Terradas-Brians\'o, S. and Masi, L. and Fattori, M. and Modugno, M.},
  journal = {Phys. Rev. Res.},
  volume = {2},
  issue = {1},
  pages = {013269},
  numpages = {8},
  year = {2020},
  month = {Mar},
  publisher = {American Physical Society},
  doi = {10.1103/PhysRevResearch.2.013269},
  url = {https://link.aps.org/doi/10.1103/PhysRevResearch.2.013269}
}

@article{Leandra,
  title = {Towards a quantum Monte Carlo--based density functional including finite-range effects: Excitation modes of a $^{39}\mathrm{K}$ quantum droplet},
  author = {Cikojevi\'c, V. and Vranje\v{s} Marki\'c, L.  and Pi, M. and Barranco, M. and Boronat, J.},
  journal = {Phys. Rev. A},
  volume = {102},
  issue = {3},
  pages = {033335},
  numpages = {8},
  year = {2020},
  month = {Sep},
  publisher = {American Physical Society},
  doi = {10.1103/PhysRevA.102.033335},
  url = {https://link.aps.org/doi/10.1103/PhysRevA.102.033335}
}

@article{Liu,
  title = {Collective excitations of a spherical ultradilute quantum droplet},
  author = {Hu, Hui and Liu, Xia-Ji},
  journal = {Phys. Rev. A},
  volume = {102},
  issue = {5},
  pages = {053303},
  numpages = {16},
  year = {2020},
  month = {Nov},
  publisher = {American Physical Society},
  doi = {10.1103/PhysRevA.102.053303},
  url = {https://link.aps.org/doi/10.1103/PhysRevA.102.053303}
}

@article{Pero,
  title = {Squeezing Bose-Bose liquid drops},
  author = {Sanuy, A. and Bara\v{c}, R. and Stipanovi\'{c}, P. and Vranje\v{s} Marki\'{c}, L.  and Boronat, J.},
  journal = {Phys. Rev. A},
  volume = {109},
  issue = {1},
  pages = {013313},
  numpages = {7},
  year = {2024},
  month = {Jan},
  publisher = {American Physical Society},
  doi = {10.1103/PhysRevA.109.013313},
  url = {https://link.aps.org/doi/10.1103/PhysRevA.109.013313}
}

@article{Boronat,
    doi = {10.1088/1367-2630/ab867a},
    url = {https://doi.org/10.1088/1367-2630/ab867a},
    year = {2020},
    month = {may},
    publisher = {IOP Publishing},
    volume = {22},
    number = {5},
    pages = {053045},
    author = {Cikojević, V and Markić, L Vranješ and Boronat, J},
    title = {Finite-range effects in ultradilute quantum drops},
    journal = {New Journal of Physics}
}

@book{Stringari,
  author = {Pitaevskii, Lev and Stringari, Sandro},
  title = {{Bose-Einstein Condensation and Superfluidity}},
  publisher = {Oxford University Press},
  year = {2016},
  month = {01},
  isbn = {9780198758884},
  DOI = {10.1093/acprof:oso/9780198758884.001.0001},
  url = {https://doi.org/10.1093/acprof:oso/9780198758884.001.0001},
}

@article{Astrakharchik,
  title = {Ultradilute Low-Dimensional Liquids},
  author = {Petrov, D. S. and Astrakharchik, G. E.},
  journal = {Phys. Rev. Lett.},
  volume = {117},
  issue = {10},
  pages = {100401},
  numpages = {5},
  year = {2016},
  month = {Sep},
  publisher = {American Physical Society},
  doi = {10.1103/PhysRevLett.117.100401},
  url = {https://link.aps.org/doi/10.1103/PhysRevLett.117.100401}
}

@article{Bohigas,
    title = {Sum rules for nuclear collective excitations},
    journal = {Physics Reports},
    volume = {51},
    number = {5},
    pages = {267-316},
    year = {1979},
    issn = {0370-1573},
    doi = {https://doi.org/10.1016/0370-1573(79)90079-6},
    url = {https://www.sciencedirect.com/science/article/pii/0370157379900796},
    author = {O. Bohigas and A.M. Lane and J. Martorell}
}

@article{bottcher_new_2020,
	title = {New states of matter with fine-tuned interactions: quantum droplets and dipolar supersolids},
	volume = {84},
	issn = {0034-4885},
	shorttitle = {New states of matter with fine-tuned interactions},
	url = {https://doi.org/10.1088/1361-6633/abc9ab},
	doi = {10.1088/1361-6633/abc9ab},
	number = {1},
	urldate = {2026-01-09},
	journal = {Reports on Progress in Physics},
	author = {Böttcher, Fabian and Schmidt, Jan-Niklas and Hertkorn, Jens and Ng, Kevin S H and Graham, Sean D and Guo, Mingyang and Langen, Tim and Pfau, Tilman},
	month = dec,
	year = {2020},
	pages = {012403},
}

@article{luo_new_2021,
	title = {A new form of liquid matter: {Quantum} droplets},
	volume = {16},
	issn = {2095-0462, 2095-0470},
	shorttitle = {A new form of liquid matter},
	url = {https://link.springer.com/10.1007/s11467-020-1020-2},
	doi = {10.1007/s11467-020-1020-2},
	number = {3},
	urldate = {2023-09-20},
	journal = {Frontiers of Physics},
	author = {Luo, Zhi-Huan and Pang, Wei and Liu, Bin and Li, Yong-Yao and Malomed, Boris A.},
	month = jun,
	year = {2021},
	pages = {32201},
}

@article{derrico_observation_2019,
  title = {Observation of quantum droplets in a heteronuclear bosonic mixture},
  author = {D'Errico, C. and Burchianti, A. and Prevedelli, M. and Salasnich, L. and Ancilotto, F. and Modugno, M. and Minardi, F. and Fort, C.},
  journal = {Phys. Rev. Res.},
  volume = {1},
  issue = {3},
  pages = {033155},
  numpages = {8},
  year = {2019},
  month = {Dec},
  publisher = {American Physical Society},
  doi = {10.1103/PhysRevResearch.1.033155},
  url = {https://link.aps.org/doi/10.1103/PhysRevResearch.1.033155}
}

@article{Cavicchioli_2025,
  title = {Dynamical Formation of Multiple Quantum Droplets in a {B}ose-{B}ose Mixture},
  author = {Cavicchioli, L. and Fort, C. and Ancilotto, F. and Modugno, M. and Minardi, F. and Burchianti, A.},
  journal = {Phys. Rev. Lett.},
  volume = {134},
  issue = {9},
  pages = {093401},
  numpages = {6},
  year = {2025},
  month = {Mar},
  publisher = {American Physical Society},
  doi = {10.1103/PhysRevLett.134.093401},
  url = {https://link.aps.org/doi/10.1103/PhysRevLett.134.093401}
}

@article{Flynn_2023,
  title = {Quantum droplets in imbalanced atomic mixtures},
  author = {Flynn, T. A. and Parisi, L. and Billam, T. P. and Parker, N. G.},
  journal = {Phys. Rev. Res.},
  volume = {5},
  issue = {3},
  pages = {033167},
  numpages = {12},
  year = {2023},
  month = {Sep},
  publisher = {American Physical Society},
  doi = {10.1103/PhysRevResearch.5.033167},
  url = {https://link.aps.org/doi/10.1103/PhysRevResearch.5.033167}
}

@article{Flynn_2024,
  title = {Harmonically trapped imbalanced quantum droplets},
  author = {Flynn, T. A. and Keepfer, N. A. and Parker, N. G. and Billam, T. P.},
  journal = {Phys. Rev. Res.},
  volume = {6},
  issue = {1},
  pages = {013209},
  numpages = {11},
  year = {2024},
  month = {Feb},
  publisher = {American Physical Society},
  doi = {10.1103/PhysRevResearch.6.013209},
  url = {https://link.aps.org/doi/10.1103/PhysRevResearch.6.013209}
}

@article{Pelayo_2025_1,
doi = {10.1088/2058-9565/ae1162},
url = {https://doi.org/10.1088/2058-9565/ae1162},
year = {2025},
month = {oct},
publisher = {IOP Publishing},
volume = {10},
number = {4},
pages = {045074},
author = {Pelayo, J C and Bougas, G A and Fogarty, T and Busch, Th and Mistakidis, S I},
title = {Droplet-gas phases and their dynamical formation in particle imbalanced mixtures},
journal = {Quantum Science and Technology}
}

@article{Petrov_2021,
  title = {Interatomic collisions in a tightly confined Bose gas},
  author = {Petrov, D. S. and Shlyapnikov, G. V.},
  journal = {Phys. Rev. A},
  volume = {64},
  issue = {1},
  pages = {012706},
  numpages = {14},
  year = {2001},
  month = {Jun},
  publisher = {American Physical Society},
  doi = {10.1103/PhysRevA.64.012706},
  url = {https://link.aps.org/doi/10.1103/PhysRevA.64.012706}
}

@article{Zin18,
  title = {Quantum {B}ose-{B}ose droplets at a dimensional crossover},
  author = {Zin, Pawe\l{} and Pylak, Maciej and Wasak, Tomasz and Gajda, Mariusz and Idziaszek, Zbigniew},
  journal = {Phys. Rev. A},
  volume = {98},
  issue = {5},
  pages = {051603},
  numpages = {5},
  year = {2018},
  month = {Nov},
  publisher = {American Physical Society},
  doi = {10.1103/PhysRevA.98.051603},
  url = {https://link.aps.org/doi/10.1103/PhysRevA.98.051603}
}

@article{Ilg,
  title = {Dimensional crossover for the beyond-mean-field correction in {B}ose gases},
  author = {Ilg, Tobias and Kumlin, Jan and Santos, Luis and Petrov, Dmitry S. and B\"uchler, Hans Peter},
  journal = {Phys. Rev. A},
  volume = {98},
  issue = {5},
  pages = {051604},
  numpages = {5},
  year = {2018},
  month = {Nov},
  publisher = {American Physical Society},
  doi = {10.1103/PhysRevA.98.051604},
  url = {https://link.aps.org/doi/10.1103/PhysRevA.98.051604}
}

@article{quasi-2D,
  title = {Ultradilute quasi-two-dimensional {B}ose-{B}ose mixtures},
  author = {Vranje\v{s} Marki\'c, Leandra and Popari\'c, Ivan and D\v{z}elalija, Kre\v{s}imir and Stipanovi\'c, Petar and Boronat, Jordi'},
  journal = {arXiv:2607.25558},
  year = {2026},
  url = {ADD}
}

@article{Ma_2023,
  title = {Quantum-fluctuation-driven dynamics of droplet splashing, recoiling, and deposition in ultracold binary {B}ose gases},
  author = {Ma, Yinfeng and Cui, Xiaoling},
  journal = {Phys. Rev. Res.},
  volume = {5},
  issue = {1},
  pages = {013100},
  numpages = {10},
  year = {2023},
  month = {Feb},
  doi = {10.1103/PhysRevResearch.5.013100},
  url = {https://link.aps.org/doi/10.1103/PhysRevResearch.5.013100}
}

@Article{Fort_2021,
AUTHOR = {Fort, Chiara and Modugno, Michele},
TITLE = {Self-Evaporation Dynamics of Quantum Droplets in a 41{K}-87{Rb} Mixture},
JOURNAL = {Applied Sciences},
VOLUME = {11},
YEAR = {2021},
NUMBER = {2},
ARTICLE-NUMBER = {866},
ISSN = {2076-3417},
}

@article{Guo_2021,
  title = {Lee-{H}uang-{Y}ang effects in the ultracold mixture of $^{23}\mathrm{Na}$ and $^{87}\mathrm{Rb}$ with attractive interspecies interactions},
  author = {Guo, Zhichao and Jia, Fan and Li, Lintao and Ma, Yinfeng and Hutson, Jeremy M. and Cui, Xiaoling and Wang, Dajun},
  journal = {Phys. Rev. Res.},
  volume = {3},
  issue = {3},
  pages = {033247},
  numpages = {8},
  year = {2021},
  month = {Sep},
  publisher = {American Physical Society},
  doi = {10.1103/PhysRevResearch.3.033247},
  url = {https://link.aps.org/doi/10.1103/PhysRevResearch.3.033247}
}

@article{pelayo_phases_2025,
	title = {Phases and dynamics of quantum droplets in the crossover to two-dimensions},
	volume = {18},
	issn = {2542-4653},
	number = {4},
	journal = {Scipost Physics},
	author = {Pelayo, Jose Carlos and Bougas, George and Fogarty, Thomas and Busch, Thomas and Mistakidis, Simeon I.},
	month = apr,
	year = {2025},
}

@article{sturmer_breathing_2021,
	title = {Breathing mode in two-dimensional binary self-bound {Bose}-gas droplets},
	volume = {103},
	url = {https://link.aps.org/doi/10.1103/PhysRevA.103.053302},
	doi = {10.1103/PhysRevA.103.053302},
	number = {5},
	urldate = {2025-06-05},
	journal = {Physical Review A},
	author = {Stürmer, P. and Tengstrand, M. Nilsson and Sachdeva, R. and Reimann, S. M.},
	month = may,
	year = {2021},
	pages = {053302},
}

@article{Stringari96,
  title = {Collective Excitations of a Trapped {B}ose-Condensed Gas},
  author = {Stringari, S.},
  journal = {Phys. Rev. Lett.},
  volume = {77},
  issue = {12},
  pages = {2360--2363},
  numpages = {0},
  year = {1996},
  month = {Sep},
  publisher = {American Physical Society},
  doi = {10.1103/PhysRevLett.77.2360},
  url = {https://link.aps.org/doi/10.1103/PhysRevLett.77.2360}
}

@article{heiselberg_collective_2004,
  title = {Collective Modes of Trapped Gases at the {BEC-BCS} Crossover},
  author = {Heiselberg, H.},
  journal = {Phys. Rev. Lett.},
  volume = {93},
  issue = {4},
  pages = {040402},
  numpages = {4},
  year = {2004},
  month = {Jul},
  publisher = {American Physical Society},
  doi = {10.1103/PhysRevLett.93.040402},
  url = {https://link.aps.org/doi/10.1103/PhysRevLett.93.040402}
}

@article{maruyama_coupled_2007,
	title = {Coupled breathing oscillations of two-component fermion condensates in deformed traps},
	volume = {75},
	url = {https://link.aps.org/doi/10.1103/PhysRevA.75.033611},
	doi = {10.1103/PhysRevA.75.033611},
	number = {3},
	urldate = {2026-07-24},
	journal = {Physical Review A},
	publisher = {American Physical Society},
	author = {Maruyama, Tomoyuki and Nishimura, Takushi},
	month = mar,
	year = {2007},
	pages = {033611},
}

@article{maruyama_deformation_2014,
	title = {Deformation dependence of breathing oscillations in {Bose}–{Fermi} mixtures at zero temperature},
	volume = {47},
	issn = {0953-4075},
	url = {https://doi.org/10.1088/0953-4075/47/22/225302},
	doi = {10.1088/0953-4075/47/22/225302},
	number = {22},
	urldate = {2026-07-26},
	journal = {Journal of Physics B: Atomic, Molecular and Optical Physics},
	publisher = {IOP Publishing},
	author = {Maruyama, Tomoyuki and Yamamoto, Tikeʼs and Nishimura, Takushi and Yabu, Hiroyuki},
	month = nov,
	year = {2014},
	pages = {225302},
}

@article{Cikojevic_KRb,
  title = {Dilute quantum liquid in a {K}-{Rb} {B}ose mixture},
  author = {Cikojevi\ifmmode \acute{c}\else \'{c}\fi{}, V. and Poli, E. and Ancilotto, F. and Vranje\v{s} Marki\'{c}, L. and Boronat, J.},
  journal = {Phys. Rev. A},
  volume = {104},
  issue = {3},
  pages = {033319},
  numpages = {10},
  year = {2021},
  month = {Sep},
  publisher = {American Physical Society},
  doi = {10.1103/PhysRevA.104.033319},
  url = {https://link.aps.org/doi/10.1103/PhysRevA.104.033319}
}

\end{document}